\documentclass[twocolumn,showpacs,preprintnumbers,amsmath,amssymb]{revtex4}
\usepackage{graphicx}
\usepackage{dcolumn}
\usepackage{bm}
\begin{document}
{\bf Comment on ``Low-Density Spin Susceptibility and Effective Mass
of Mobile Electrons in Si Inversion Layers''}
\vspace{.15in}

In a recent Letter \cite{pudalov02}, the effective $g$-factor and the
effective mass, $m$, have been studied by Pudalov {\it et al.} in a
dilute 2D electron system in silicon. By analyzing Shubnikov-de~Haas
oscillations in superimposed parallel and perpendicular magnetic
fields, the authors reproduce the strong increase of $gm$ with
decreasing electron density previously reported in
Ref.~\cite{kravchenko00}. However, they contrast their data with the
data obtained by our group \cite{shashkin01} and claim that the spin
susceptibility (or the product $gm$) ``increases gradually with
decreasing density'', which ``does not support the occurrence of
spontaneous spin polarization and divergence of $gm$ at $n_s=n_c$''
(here $n_s$ is the electron density and $n_c$ is the critical density
for the metal-insulator transition, MIT). The purpose of this Comment
is to show that all available experimental data, including those of
Pudalov {\it et al.}, are consistent with each other and are in favor
of a spontaneous spin polarization in this 2D system and a divergence
of $gm$ at a finite electron density.

Spin polarization and spin susceptibility were recently studied by
measuring the (parallel) magnetic field $B_c$, required to fully
polarize the electrons' spins, using scaling of magnetoresistance
\cite{shashkin01} and magnetoconductivity \cite{vitkalov01}. To
compare the data of the above two groups with the newer data of
Pudalov {\it et al.} \cite{pudalov02}, in Fig.~1 we plot all three
sets of data (two sets in the inset for better visibility). To
convert $gm$ from Ref.~\cite{pudalov02} into $B_c$, we use the
condition for the full spin polarization:
$g\mu_BB_c/2=\pi\hbar^2n_s/2m$, where $\mu_B$ is the Bohr magneton.
(The factor of 2 on the right side of the equation reflects the
valley degeneracy.) The agreement between all three sets of data is
remarkable, especially if one takes into account that different
groups used different methods, different samples, and different
field/spin-polarization ranges. $B_c$ is a linear function of $n_s$
and extrapolates to zero at a {\it finite} electron density which we
will designate $n_\chi$. The linear fit of the data from
Ref.~\cite{pudalov02} yields $n_\chi=8\times10^{10}$~cm$^{-2}$, which
is identical with ours suggesting that $n_\chi$ is
sample-independent.

Contrary to the claim made in Ref.~\cite{pudalov02} of a {\em
gradual} increase of $\chi$ with decreasing $n_s$, the linear
dependence of $B_c\propto n_s/gm$ (Fig.~1) points to the {\em
critical} behavior of the spin susceptibility: $\chi\propto
n_s/(n_s-n_\chi)$. The divergence of $\chi$ should occur at the
sample-independent electron density $n_\chi$, which in the samples
studied in Ref.~\cite{shashkin01} coincides with the critical density
$n_c$ for the MIT indicated by arrow. In more disordered samples,
however, $n_c$ may be noticeably higher than $n_\chi$. This is the
case for the samples from Ref.~\cite{pudalov02} with
$n_c\approx1\times10^{11}$~cm$^{-2}$ being well above the expected
ferromagnetic transition point
$n_\chi\approx8\times10^{10}$~cm$^{-2}$. This explains why no
divergence of $gm$ is seen by Pudalov {\it et al.} at electron
densities down to $n_c$ in their samples. Note that even in the least
disordered samples $gm$ is still expected to be finite near $n_\chi$,
as it normally occurs for any ferromagnetic transition due to
non-zero temperature, inhomogeneous broadening, {\it etc}.

Of course, for the spin susceptibility to diverge at $n_s=n_\chi$,
the extrapolation of $B_c(n_s)$ to zero must be valid. To verify its
validity, accurate data at lower densities, lower temperatures, and
on much less disordered samples are needed. We emphasize that, in
contrast to their claim, the method used by Pudalov {\it et al.}
\cite{pudalov02} certainly cannot be applied ``down to and across the
2D MIT'' because at $n_s\lesssim10^{11}$~cm$^{-2}$, (i)~the amplitude
of oscillations is too large (and even diverges as $T\rightarrow0$)
\cite{pudalov94}, which is inconsistent with the Lifshitz-Kosevich
formula they use, and (ii)~there are too few oscillations
\cite{pudalov94} to study the beating pattern. We also note that the
Lifshitz-Kosevich formula was deduced for the case of weak
electron-electron interactions, and its application to
strongly-correlated system is not justified.

\begin{figure}\vspace{2mm}
\scalebox{0.4}{\includegraphics{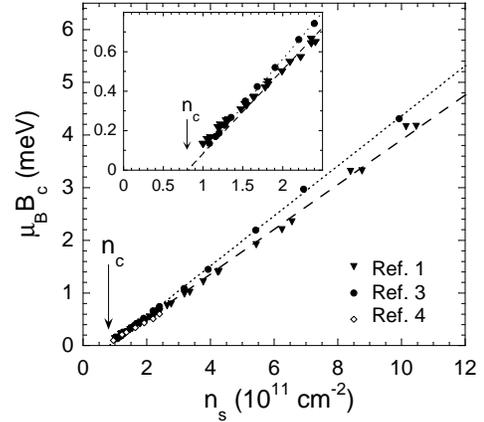}}\vspace{-5cm}
\caption{\label{fig.1} $B_c(n_s)$ calculated using the data from
Refs.~\protect\cite{pudalov02,shashkin01,vitkalov01}. The dashed and
dotted straight lines are fits to the data from
Refs.~\protect\cite{pudalov02,shashkin01},
respectively.}\vspace{-5mm}
\end{figure}

\noindent S.~V. Kravchenko\\
{\small\indent Physics Department\\
\indent Northeastern University\\
\indent Boston, Massachusetts 02115}

\noindent A.~A. Shashkin and V.~T. Dolgopolov\\
{\small\indent Institute of Solid State Physics\\
\indent Chernogolovka, Moscow District 142432\\
\indent Russia}

\end{document}